\documentclass[sigconf]{acmart}

\usepackage{booktabs}
\usepackage{graphicx, epstopdf}
\usepackage{subfigure}
\usepackage{url}
\usepackage{multirow}
\usepackage{algorithm}
\usepackage{algorithmic}
\usepackage{balance}
\usepackage{arydshln}

\begin{document}

\title{MiNet: Mixed Interest Network for Cross-Domain Click-Through Rate Prediction}
\author{Wentao Ouyang, \hskip -0.8pt Xiuwu Zhang, \hskip -0.8pt Lei Zhao, \hskip -0.8pt Jinmei Luo, \hskip -0.8pt Yu Zhang, \hskip -0.8pt Heng Zou, \hskip -0.8pt Zhaojie Liu, \hskip -0.8pt Yanlong Du}
\affiliation{%
  \institution{Intelligent Marketing Platform, Alibaba Group}
}
\email{{maiwei.oywt, xiuwu.zxw, zhaolei.zl, cathy.jm, zy107620, zouheng.zh, zhaojie.lzj, yanlong.dyl}@alibaba-inc.com}

\begin{abstract}
Click-through rate (CTR) prediction is a critical task in online advertising systems. Existing works mainly address the \emph{single-domain} CTR prediction problem and model aspects such as feature interaction, user behavior history and contextual information. Nevertheless, ads are usually displayed with natural content, which offers an opportunity for \emph{cross-domain} CTR prediction. In this paper, we address this problem and leverage auxiliary data from a source domain to improve the CTR prediction performance of a target domain. Our study is based on UC Toutiao (a news feed service integrated with the UC Browser App, serving hundreds of millions of users daily), where the source domain is the news and the target domain is the ad.
In order to effectively leverage news data for predicting CTRs of ads, we propose the Mixed Interest Network (MiNet) which jointly models three types of user interest: 1) long-term interest across domains, 2) short-term interest from the source domain and 3) short-term interest in the target domain. MiNet contains two levels of attentions, where the \emph{item-level} attention can adaptively distill useful information from clicked news / ads and the \emph{interest-level} attention can adaptively fuse different interest representations. Offline experiments show that MiNet outperforms several state-of-the-art methods for CTR prediction. We have deployed MiNet in UC Toutiao and the A/B test results show that the online CTR is also improved substantially. MiNet now serves the main ad traffic in UC Toutiao.
\end{abstract}

\ccsdesc[500]{Information systems~Online advertising}
\ccsdesc[500]{Information systems~Computational advertising}

\keywords{Click-through rate prediction; Online advertising; Deep learning}

\copyrightyear{2020}
\acmYear{2020}
\setcopyright{acmcopyright}\acmConference[CIKM '20]{Proceedings of the 29th ACM International Conference on Information and Knowledge Management}{October 19--23, 2020}{Virtual Event, Ireland}
\acmBooktitle{Proceedings of the 29th ACM International Conference on Information and Knowledge Management (CIKM '20), October 19--23, 2020, Virtual Event, Ireland}
\acmPrice{15.00}
\acmDOI{10.1145/3340531.3412728}
\acmISBN{978-1-4503-6859-9/20/10}

\settopmatter{printacmref=true}
\fancyhead{}

\maketitle

\section{Introduction}
Click-through rate (CTR) prediction plays an important role in online advertising systems. It aims to predict the probability that a user will click on a specific ad. The predicted CTR impacts both the ad ranking strategy and the ad charging model \cite{zhou2018deep,ouyang2019deep}.
Therefore, in order to maintain a desirable user experience and to maximize the revenue, it is crucial to estimate the CTR of ads accurately.

CTR prediction has attracted lots of attention from both academia and industry \cite{he2014practical,cheng2016wide,shan2016deep,zhang2016deep,he2017neural,zhou2018deep}. For example, Factorization Machine (FM) \cite{rendle2010factorization} is proposed to model pairwise feature interactions. Deep Neural Networks (DNNs) are exploited for CTR prediction and item recommendation in order to automatically learn feature representations and high-order feature interactions \cite{van2013deep,zhang2016deep,covington2016deep}. To take advantage of both shallow and deep models, hybrid models such as Wide\&Deep \cite{cheng2016wide} (which combines Logistic Regression and DNN) are also proposed.
Moreover, Deep Interest Network (DIN) \cite{zhou2018deep} models dynamic user interest based on historical behavior. Deep Spatio-Temporal Network (DSTN) \cite{ouyang2019deep} jointly exploits contextual ads, clicked ads and unclicked ads for CTR prediction.

\begin{figure}[!t]
\centering
\includegraphics[width=0.44\textwidth, trim = 0 0 0 0, clip]{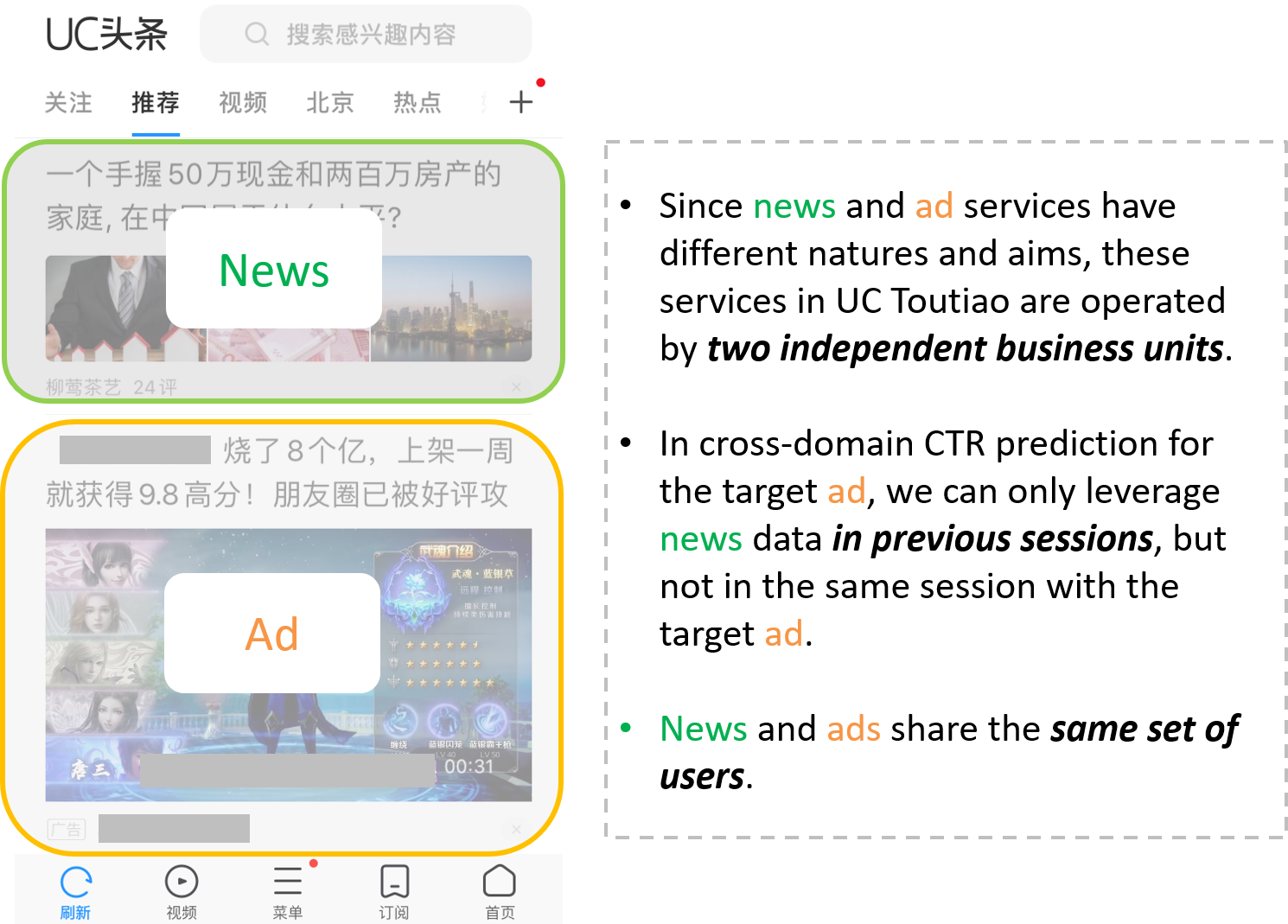}
\vskip -6pt
\caption{Illustration of news and ads in UC Toutiao.}
\vskip -8pt
\label{illus}
\end{figure}

As can be seen, existing works mainly address \emph{single-domain} CTR prediction, i.e., they only utilize ad data for CTR prediction and they model aspects such as feature interaction \cite{rendle2010factorization}, user behavior history \cite{zhou2018deep,ouyang2019deep} and contextual information \cite{ouyang2019deep}.
Nevertheless, ads are usually displayed with natural content, which offers an opportunity for \emph{cross-domain} CTR prediction. In this paper, we address this problem and leverage auxiliary data from a source domain to improve the CTR prediction performance of a target domain.
Our study is based on UC Toutiao (Figure \ref{illus}), where the source domain is the natural news feed (news domain) and the target domain is the advertisement (ad domain).
\emph{A major advantage} of cross-domain CTR prediction is that by enriching data across domains, the data sparsity and the cold-start problem in the target domain can be alleviated, which leads to improved prediction performance.

In order to effectively leverage cross-domain data, we consider three types of user interest as follows:
\begin{itemize}
\item \textbf{Long-term interest across domains.}
Each user has her own profile features such as user ID, age group, gender and city. These profile features reflect a user's long-term intrinsic interest. Based on cross-domain data (i.e., all the news and ads that the user has interacted with), we are able to learn more semantically richer and more statistically reliable user feature embeddings.

\item \textbf{Short-term interest from the source domain}.
For each target ad whose CTR is to be predicted, there is corresponding short-term user behavior in the source domain (e.g., news the user just viewed). Although the content of a piece of news can be completely different from that of the target ad, there may exist certain correlation between them. For example, a user has a high probability to click a \emph{Game ad} after viewing some \emph{Entertainment news}. Based on such relationships, we can transfer useful knowledge from the source domain to the target domain.

\item \textbf{Short-term interest in the target domain}.
For each target ad, there is also corresponding short-term user behavior in the target domain. What ads the user has recently clicked may have a strong implication for what ads the user may click in the short future.
\end{itemize}

Although the above proposal looks promising, it faces several challenges. 1) Not all the pieces of clicked news are indicative of the CTR of the target ad. 2) Similarly, not all clicked ads are informative about the CTR of the target ad. 3) The model must be able to transfer knowledge from news to ads. 4) The relative importance of the three types of user interest may vary w.r.t. different target ads. For example, if the target ad is similar to a recently clicked ad, then the short-term interest in the target domain should be more important; if the target ad is irrelevant to both the recently clicked news and ads, then the long-term interest should be more important. 5) The representation of the target ad and those of the three types of user interest may have different dimensions (due to different numbers of features). The dimension discrepancy may naturally boost or weaken the impact of some representations, which is undesired.

To address these challenges, we propose the Mixed Interest Network (MiNet), whose structure is shown in Figure \ref{model_minet}.
In MiNet, the long-term user interest is modeled by the concatenation of user profile feature embeddings $\mathbf{p}_u$, which is jointly learned based on cross-domain data, enabling knowledge transfer; the short-term interest from the source domain is modeled by the vector $\mathbf{a}_s$, which aggregates the information of recently clicked news; the short-term interest in the target domain is modeled by the vector $\mathbf{a}_t$, which aggregates the information of recently clicked ads.

MiNet contains two levels of attentions (i.e., item-level and interest-level).
The item-level attention is applied to both the source domain and the target domain, which can adaptively distill useful information from recently clicked news / ads (to tackle Challenges 1 and 2). A transfer matrix is introduced to transfer knowledge from news to ads (to tackle Challenge 3). Moreover, the long-term user interest is learned based on cross-domain data, also enabling knowledge transfer (to tackle Challenge 3).
We further introduce the interest-level attention to dynamically adjust the importance of the three types of user interest w.r.t. different target ads (to tackle Challenge 4). The interest-level attention with a proper activation function can also handle the dimension discrepancy issue (to tackle Challenge 5). Both offline and online experimental results demonstrate the effectiveness of MiNet for more accurate CTR prediction.

The main contributions of this work are summarized as follows:
\begin{enumerate}
\item We propose to jointly consider three types of user interest for cross-domain CTR prediction: 1) long-term interest across domains, 2) short-term interest from the source domain, and 3) short-term interest in the target domain.
\item We propose the MiNet model to achieve the above goal. MiNet contains two levels of attentions, where the item-level attention can adaptively distill useful information from clicked news / ads and the interest-level attention can adaptively fuse different interest representations. We make the implementation code publicly available\footnote{https://github.com/oywtece/minet}.
\item We conduct extensive offline experiments to test the performance of MiNet and several state-of-the-art methods for CTR prediction. We also conduct ablation studies to provide further insights behind the model.
\item We have deployed MiNet in UC Toutiao and conducted online A/B test to evaluate its performance in real-world CTR prediction tasks.
\end{enumerate}

\begin{table}[!t]
\caption{Each row is an instance for CTR prediction. The first column is the label (1 - clicked, 0 - unclicked). Each of the other columns is a field. Instantiation of a field is a feature.}
\vskip -8pt
\label{tab_ft}
\centering
\begin{tabular}{|c|c|c|c|c|}
\hline
\textbf{Label} & \textbf{User ID} & \textbf{User Age} & \textbf{Ad Title} \\
\hline
1 & 2135147 & 24 & Beijing flower delivery \\
\hline
0 & 3467291 & 31 & Nike shoes, sporting shoes \\
\hline
0 & 1739086 & 45 & Female clothing and jeans \\
\hline
\end{tabular}
\vskip -8pt
\end{table}

\section{Model Design}

We first introduce some notations used in the paper. We represent matrices, vectors and scalars as bold capital letters (e.g., $\mathbf{X}$), bold lowercase letters (e.g., $\mathbf{x}$) and normal lowercase letters (e.g., $x$), respectively.
By default, all vectors are in a column form.

\subsection{Problem Formulation}
The task of CTR prediction in online advertising is to build a prediction model to estimate the probability of a user clicking on a specific ad.
Each instance can be described by multiple \emph{fields} such as user information (``User ID'', ``City'', ``Age'', etc.) and ad information (``Creative ID'', ``Campaign ID'', ``Title'', etc.). The instantiation of a field is a \emph{feature}. For example, the ``User ID'' field may contain features such as ``2135147'' and ``3467291''. Table \ref{tab_ft} shows some examples.

\begin{figure}[!t]
\centering
\includegraphics[width=0.5\textwidth, trim = 0 0 0 0, clip]{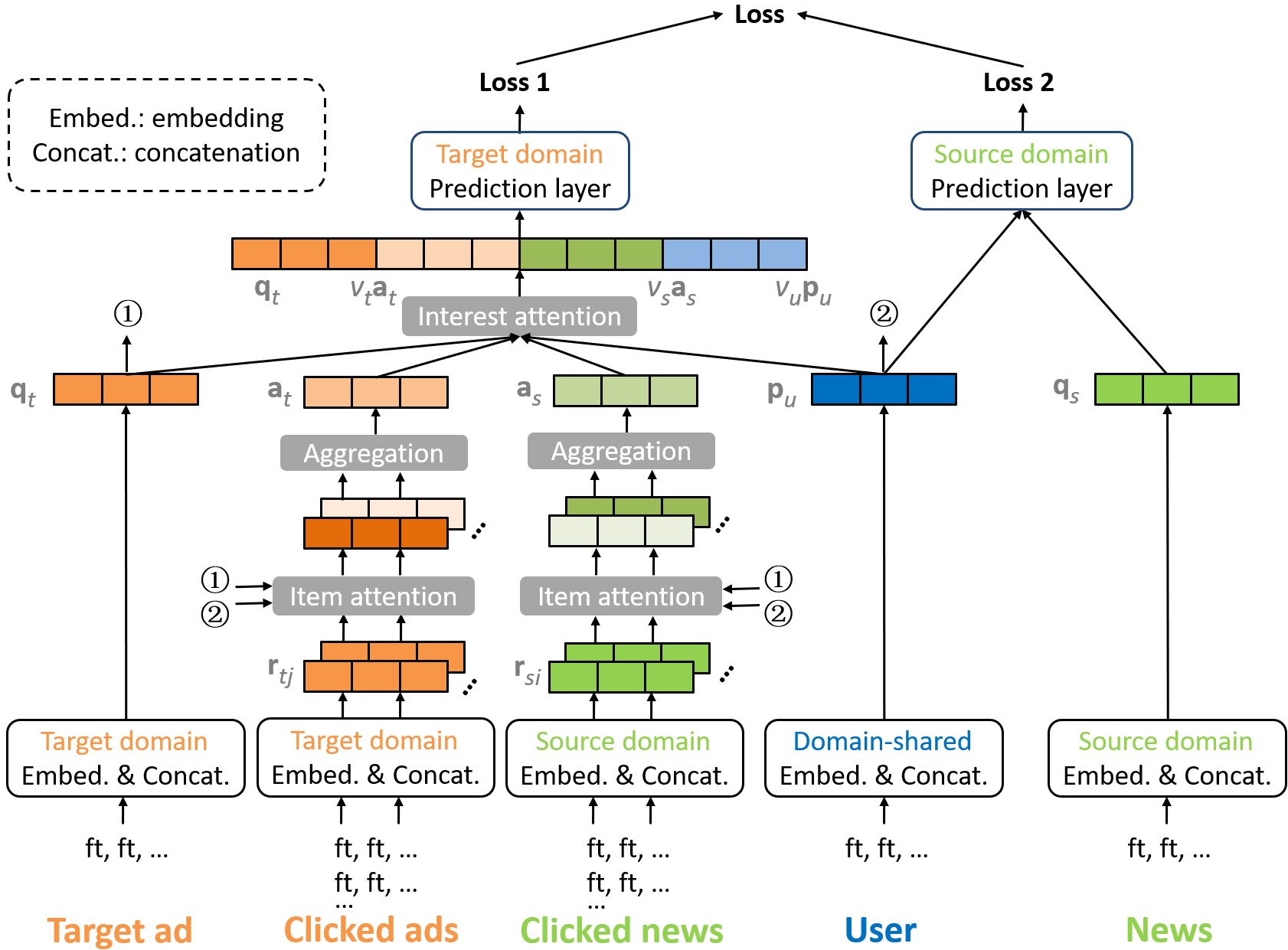}
\vskip -6pt
\caption{Structure of Mixed Interest Network (MiNet). MiNet models three types of user interest: long-term interest across domains $\mathbf{p}_u$, short-term interest from the source domain $\mathbf{a}_s$ and short-term interest in the target domain $\mathbf{a}_t$.}
\vskip -8pt
\label{model_minet}
\end{figure}

We define the cross-domain CTR prediction problem as leveraging the data from a \emph{source} domain (or source domains) to improve the CTR prediction performance in a \emph{target} domain.
In news feed advertising (e.g., UC Toutiao shown in Figure \ref{illus}), the source domain is the natural news feed and the target domain is the advertisement.
In this scenario, the source domain and the target domain \emph{share the same set of users}, but there are \emph{no overlapped items}.

\subsection{Model Overview}
In order to effectively leverage cross-domain data, we propose the Mixed Interest Network (MiNet) in Figure \ref{model_minet}, which models three types of user interest: 1) \textbf{Long-term interest across domains}. It is modeled by the concatenation of user profile feature embeddings $\mathbf{p}_u$, which is jointly learned based on cross-domain data. 2) \textbf{Short-term interest from the source domain}. It is modeled by the vector $\mathbf{a}_s$, which aggregates the information of recently clicked news in the source domain. 3) \textbf{Short-term interest in the target domain}. It is modeled by the vector $\mathbf{a}_t$, which aggregates the information of recently clicked ads in the target domain.

In MiNet, we apply two levels of attentions, one on the item level and the other on the interest level. The aim of the item-level attention is to dynamically distill useful information (relevant to the target ad whose CTR is to be predicted) from recently clicked news / ads and suppress noise. The aim of the interest-level attention is to adaptively adjust the importance of the three types of user interest (i.e., $\mathbf{p}_u$, $\mathbf{a}_s$ and $\mathbf{a}_t$) and to emphasize more informative signal(s) for the target ad. In the following, we detail our model components.

\subsection{Feature Embedding}
We first encode features into one-hot encodings. For the $i$th feature, its one-hot encoding is $\mathbf{v}_i = \textrm{one-hot}(i)$,
where $\mathbf{v}_i \in \mathbb{R}^N$ is a vector with 1 at the $i$th entry and 0 elsewhere, and $N$ is the number of unique features.
To enable cross-domain knowledge transfer, the set of unique features includes all the features in both domains.

We then map sparse, high-dimensional one-hot encodings to dense, low-dimensional embedding vectors suitable for neural networks \cite{mikolov2013distributed}.
In particular, we define an embedding matrix $\mathbf{E} \in \mathbb{R}^{D \times N}$ (to be learned; $D$ is the embedding dimension; $D \ll N$). The $i$th feature is then projected to its corresponding embedding vector $\mathbf{e}_i \in \mathbb{R}^D$ as
$\mathbf{e}_i = \mathbf{E} \mathbf{v}_i$.
Equivalently, the embedding vector $\mathbf{e}_i$ is the $i$th column of the embedding matrix $\mathbf{E}$.

\subsection{Long-term Interest across Domains}
For each ad instance, we split its features into user features and ad features. We take out all ad features and concatenate the corresponding feature embedding vectors to obtain an ad representation vector $\mathbf{r}_t \in \mathbb{R}^{D_t}$ in the target domain. Similarly, we can obtain a news representation vector $\mathbf{r}_s \in \mathbb{R}^{D_s}$ in the source domain.

For a user $u$, her long-term interest representation vector $\mathbf{p}_u \in \mathbb{R}^{D_u}$ is obtained by concatenating the corresponding user feature embedding vectors ($D_t$, $D_s$ and $D_u$ are data-related dimensions).
For example, if user $u$ has features ``UID = u123, City = BJ, Gender = male, OS = ios'', we have
\[
\mathbf{p}_u = [\mathbf{e}_\textrm{u123} \| \mathbf{e}_\textrm{BJ} \|  \mathbf{e}_\textrm{male} \| \mathbf{e}_\textrm{ios}],
\]
where $\|$ is the vector concatenation operation.

This long-term interest representation vector $\mathbf{p}_u$ is shared across domains and is jointly learned using the data from both domains. The detailed process will be described in \S\ref{prediction} and \S\ref{learning}.

\subsection{\!\!Short-term Interest from the Source Domain}
Given a user, for each target ad whose CTR is to be predicted in the target domain, the user usually viewed pieces of news in the source domain. Although the content of a piece of news can be completely different from that of the target ad, there may exist certain correlation between them. For example, a user has a high probability to click a Game ad after viewing some Entertainment news. Based on such relationships, we can transfer useful knowledge from the source domain to the target domain.

Denote the set of representation vectors of recently clicked news as $\{\mathbf{r}_{si}\}_i$.
Because the number of pieces of clicked news may be different from time to time, we need to aggregate these pieces of news.
In particular, the aggregated representation $\mathbf{a}_s$ is given by
\begin{equation} \label{int_src}
\mathbf{a}_s = \sum_i \alpha_i \mathbf{r}_{si},
\end{equation}
where $\alpha_i$ is a weight assigned to $\mathbf{r}_{si}$ to indicate its importance during aggregation.
The aggregated representation $\mathbf{a}_s$ reflects the short-term interest of the user from the source domain.

The problem remaining is how to compute the weight $\alpha_i$. One simple way is to set $\alpha_i = 1/|\{\mathbf{r}_{si}\}_i|$.
That is, each piece of clicked news has equal importance. This is clearly not a wise choice because some news may not be indicative for the target ad.

Alternatively, the attention mechanism \cite{bahdanau2014neural} offers us a better way of computing $\alpha_i$.
It is firstly introduced in the encoder-decoder framework for the machine translation task.
Nevertheless, how to use it is still flexible. One possible way is to compute $\alpha_i$ as
\begin{equation}
\alpha_i = \frac{\exp(\tilde{\alpha}_i)}{ \sum_{i'} \exp(\tilde{\alpha}_{i'})},
\end{equation}
\begin{equation} \label{att_simple}
\tilde{\alpha}_i = \mathbf{h}_s ^T ReLU\big(\mathbf{W}_s \mathbf{r}_{si} \big),
\end{equation}
where $\mathbf{W}_s$ and $\mathbf{h}_s$ are model parameters; $ReLU$ is the rectified linear unit ($ReLU(x) = \max(0, x)$) as an activation function. Nair and Hinton \cite{nair2010rectified} show that ReLU has significant benefits over sigmoid and tanh activation functions in terms of the convergence rate and the quality of obtained results.

However, Eq. (\ref{att_simple}) only considers each piece of clicked news $\mathbf{r}_{si}$ alone. It does not capture the relationship between a piece of clicked news and the target ad. Moreover, Eq. (\ref{att_simple}) is not tailored to the target user as well.
For example, no matter the target ad is about coffee or clothing, or the target user is $u_a$ or $u_b$, the importance of a piece of clicked news keeps the same.

\subsubsection{Item-level Attention}
Given the above limitations, we actually compute $\tilde{\alpha}_i$ as
\begin{equation} \label{att_src}
\tilde{\alpha}_i = \mathbf{h}_s ^T ReLU\big(\mathbf{W}_s [\mathbf{r}_{si} \| \mathbf{q}_t \| \mathbf{p}_u \| \mathbf{M} \mathbf{r}_{si} \odot \mathbf{q}_t] \big),
\end{equation}
where $\mathbf{W}_s \in \mathbb{R}^{D_h \times (D_s + 2D_t + D_u) }$, $\mathbf{h}_s \in \mathbb{R}^{D_h}$, $\mathbf{M} \in \mathbb{R}^{D_t \times D_s}$ are parameters to be learned ($D_h$ is a dimension hyperparameter).

Eq. (\ref{att_src}) considers the following aspects:
\begin{itemize}
\item The \textbf{clicked news} $\mathbf{r}_{si} \in \mathbb{R}^{D_s}$ in the source domain.
\item The \textbf{target ad} $\mathbf{q}_t \in \mathbb{R}^{D_t}$ in the target domain.
\item The \textbf{target user} $\mathbf{p}_u \in \mathbb{R}^{D_u}$.
\item The \textbf{transferred interaction} $\mathbf{M} \mathbf{r}_{si} \odot \mathbf{q}_t \in \mathbb{R}^{D_t}$ between the clicked news and the target ad. $\odot$ is the element-wise product operator. $\mathbf{M}$ is a transfer matrix that transfers $\mathbf{r}_{si} \in \mathbb{R}^{D_s}$ in the source domain to $\mathbf{M} \mathbf{r}_{si} \in \mathbb{R}^{D_t}$ in the target domain such that $\mathbf{M} \mathbf{r}_{si}$ can be compared with the target ad $\mathbf{q}_t$.
\end{itemize}
In this way, the computed $\tilde{\alpha}_i$ ($\alpha_i$) is not only a function of the clicked news which needs to be assigned a weight, but is also aware of the target ad and the target user. It also considers the interaction between the clicked news and the target ad across domains.

\subsubsection{Complexity Reduction}
In Eq. (\ref{att_src}), the transfer matrix $\mathbf{M}$ is of dimension $D_t \times D_s$. When $D_t$ and $D_s$ are large, $\mathbf{M}$ contains lots of parameters to be learned. To reduce the computational complexity, we decompose $\mathbf{M}$ as
$\mathbf{M} = \mathbf{M}_1 \times \mathbf{M}_2$,
where $\mathbf{M}_1 \in \mathbb{R}^{D_t \times C}$ and $\mathbf{M}_2 \in \mathbb{R}^{C \times D_s}$. $C$ is an intermediate dimension, which can be set to a small number.
In this way, the total number of parameters reduces from $D_t \times D_s$ to $(D_t + D_s) \times C$.

\subsection{Short-term Interest in the Target Domain}
Given a user, for each target ad whose CTR is to be predicted, the user also has recent behavior in the target domain. What ads the user has recently clicked may have a strong implication for what ads the user may click in the short future.

Denote the set of representation vectors of recently clicked ads as $\{\mathbf{r}_{tj}\}_j$.
We compute the aggregated representation $\mathbf{a}_t$ as
\begin{equation} \label{int_tar}
\mathbf{a}_t = \sum_j \beta_j \mathbf{r}_{tj}, \ \beta_j = \frac{\exp(\tilde{\beta}_j)}{ \sum_{j'} \exp(\tilde{\beta}_{j'})},
\end{equation}
\begin{equation} \label{att_tar}
\tilde{\beta}_j = \mathbf{h}_t ^T ReLU\big(\mathbf{W}_t [\mathbf{r}_{tj} \| \mathbf{q}_t \| \mathbf{p}_u \| \mathbf{r}_{tj} \odot \mathbf{q}_t] \big),
\end{equation}
where $\mathbf{W}_t \in \mathbb{R}^{D_h \times (3D_t + D_u)}$ and $\mathbf{h}_t \in \mathbb{R}^{D_h}$ are parameters to be learned.
The representation $\mathbf{a}_t$ reflects the short-term interest of the user in the target domain.

Eq. (\ref{att_tar}) considers the following aspects:
\begin{itemize}
\item The \textbf{clicked ad} $\mathbf{r}_{tj} \in \mathbb{R}^{D_t}$ in the target domain.
\item The \textbf{target ad} $\mathbf{q}_t \in \mathbb{R}^{D_t}$ in the target domain.
\item The \textbf{target user} $\mathbf{p}_u \in \mathbb{R}^{D_u}$.
\item The \textbf{interaction} $\mathbf{r}_{tj} \odot \mathbf{q}_t \in \mathbb{R}^{D_t}$ between the clicked ad and the target ad. Because they are in the same domain, no transfer matrix is needed.
\end{itemize}
Similarly, the computed $\tilde{\beta}_j$ ($\beta_j$) is not only a function of the clicked ad which needs to be assigned a weight, but is also aware of the target ad and the target user.

\subsection{Interest-Level Attention}
After we obtain the three types of user interest $\mathbf{p}_u \in \mathbb{R}^{D_u}$, $\mathbf{a}_s \in \mathbb{R}^{D_s}$ and $\mathbf{a}_t \in \mathbb{R}^{D_t}$, we use them together to predict the CTR of the target ad $\mathbf{q}_t \in \mathbb{R}^{D_t}$.
Although $\mathbf{p}_u$, $\mathbf{a}_s$ and $\mathbf{a}_t$ all represent user interest, they reflect different aspects and have different dimensions. We thus cannot use weighted sum to fuse them.
One possible solution is to concatenate all available information as a long input vector
\begin{equation}
\mathbf{m} \triangleq [\mathbf{q}_t \| \mathbf{p}_u \| \mathbf{a}_s \| \mathbf{a}_t].
\end{equation}
However, such a solution cannot find the most informative user interest signal for the target ad $\mathbf{q}_t$. For example, if both the short-term interest $\mathbf{a}_s$ and $\mathbf{a}_t$ are irrelevant to the target ad $\mathbf{q}_t$, the long-term interest $\mathbf{p}_u$ should be more informative. But $\mathbf{p}_u$, $\mathbf{a}_s$ and $\mathbf{a}_t$ have equal importance in $\mathbf{m}$.

Therefore, instead of forming $\mathbf{m}$, we actually form $\mathbf{m}_t$ as follows
\begin{equation} \label{m_t}
\mathbf{m}_t \triangleq [\mathbf{q}_t \| v_u \mathbf{p}_u \| v_s \mathbf{a}_s \| v_t \mathbf{a}_t],
\end{equation}
where $v_u$, $v_s$ and $v_t$ are dynamic weights that tune the importance of different user interest signals based on their actual values.

In particular, we compute these weights as follows:
\begin{align} \label{att_int}
v_u &= \exp \left(\mathbf{g}_u^T ReLU\big( \mathbf{V}_u [\mathbf{q}_t \| \mathbf{p}_u \| \mathbf{a}_s \| \mathbf{a}_t]\big)  + b_u \right), \nonumber \\
v_s &= \exp \left(\mathbf{g}_s^T ReLU\big( \mathbf{V}_s [\mathbf{q}_t \| \mathbf{p}_u \| \mathbf{a}_s \| \mathbf{a}_t]\big)  + b_s \right), \nonumber \\
v_t &= \exp \left(\mathbf{g}_t^T ReLU\big( \mathbf{V}_t [\mathbf{q}_t \| \mathbf{p}_u \| \mathbf{a}_s \| \mathbf{a}_t]\big)  + b_t \right),
\end{align}
where $\mathbf{V}_* \in \mathbb{R}^{D_h \times (D_s + 2D_t + D_u) }$ is a matrix parameter, $\mathbf{g}_* \in \mathbb{R}^{D_h}$ is a vector parameter and $b_*$ is a scalar parameter. The introduction of $b_*$ is to model the intrinsic importance of a particular type of user interest, regardless of its actual value.
It is observed that these weights are computed based on all the available information so as to take into account the contribution of a particular type of user interest to the target ad, given other types of user interest signals.

It is also observed that we use $\exp(\cdot)$ to compute the weights, which makes $v_*$ may be larger than 1. It is a desirable property because these weights can compensate for the dimension discrepancy problem. For example, when the dimension of $\mathbf{q}_t$ is much larger than that of $\mathbf{p}_u$ (due to more features), the contribution of $\mathbf{p}_u$ would be naturally weakened by this effect. Assigning $\mathbf{p}_u$ with a weight in $[0, 1]$ (i.e., replacing $\exp(\cdot)$ by the sigmoid function) cannot address this issue.
Nevertheless, as these weights are automatically learned, they could be smaller than 1 as well when necessary.

\subsection{Prediction} \label{prediction}
In the target domain, we let the input vector $\mathbf{m}_t$ go through several fully connected (FC) layers with the ReLU activation function, in order to exploit high-order feature interaction as well as nonlinear transformation \cite{he2017neural}.
Formally, the FC layers are defined as follows:
\begin{align}
\mathbf{z}_1 & = ReLU(\mathbf{W}_1 \mathbf{m}_t + \mathbf{b}_1), \
\mathbf{z}_2 = ReLU(\mathbf{W}_2 \mathbf{z}_1 + \mathbf{b}_2), \ \cdots \nonumber \\
\mathbf{z}_L & = ReLU(\mathbf{W}_L \mathbf{z}_{L-1} + \mathbf{b}_L),  \nonumber
\end{align}
where $L$ denotes the number of hidden layers; $\mathbf{W}_l$ and $\mathbf{b}_l$ denote the weight matrix and bias vector (to be learned) in the $l$th layer.

Finally, the vector $\mathbf{z}_L$ goes through an output layer with the sigmoid function to generate the predicted CTR of the target ad as
\[
\hat{y}_t = \frac{1}{1+\exp[- (\mathbf{w}^T \mathbf{z}_L + b)]},
\]
where $\mathbf{w}$ and $b$ are the weight and bias parameters to be learned.

To facilitate the learning of long-term user interest $\mathbf{p}_u$, we also create an input vector for the source domain as
$\mathbf{m}_s \triangleq [\mathbf{q}_s \| \mathbf{p}_u]$,
where $\mathbf{q}_s \in \mathbb{R}^{D_s}$ is the concatenation of the feature embedding vectors for the target news. Similarly, we let $\mathbf{m}_s$ go through several FC layers and an output layer (with their own parameters). Finally, we obtain the predicted CTR $\hat{y}_s$ of the target news.

\subsection{Model Learning} \label{learning}
We use the cross-entropy loss as our loss function.
In the target domain, the loss function on a training set is given by
\begin{equation} \label{loss_t}
loss_t = - \frac{1}{|\mathbb{Y}_t|}\sum_{y_t \in \mathbb{Y}_t} [y_t \log \hat{y}_t + (1 - y_t) \log (1 - \hat{y}_t)],
\end{equation}
where $y_t \in\{0,1\}$ is the true label of the target ad corresponding to the estimated CTR $\hat{y}_t$ and $\mathbb{Y}_t$ is the collection of true labels.
Similarly, we have a loss function $loss_s$ in the source domain.

All the model parameters are learned by minimizing the combined loss as follows
\begin{equation}
loss = loss_t + \gamma loss_s,
\end{equation}
where $\gamma$ is a balancing hyperparameter.
As $\mathbf{p}_u$ is shared across domains, when optimizing the combined loss, $\mathbf{p}_u$ is jointly learned based on the data from both domains.

\section{Experiments}
In this section, we conduct both offline and online experiments to evaluate the performance of the proposed MiNet as well as several state-of-the-art methods for CTR prediction.

\subsection{Datasets}
The statistics of the datasets are listed in Table \ref{tab_stat}.

1) \textbf{Company News-Ads dataset.}
This dataset contains a random sample of news and ads impression and click logs from the news system and the ad system in UC Toutiao.
The source domain is the news and the target domain is the ad.
We use logs of 6 consecutive days in 2019 for initial training, logs of the next day for validation, and logs of the day after the next day for testing. After finding the optimal hyperparameters on the validation set, we combine the initial training set and the validation set as the final training set (trained using the found optimal hyperparameters).
The features used include 1) user features such as user ID, agent and city, 2) news features such as news title, category and tags, and 3) ad features such as ad title, ad ID and cateogry.

2) \textbf{Amazon Books-Movies dataset.}
The Amazon datasets \cite{mcauley2015image} have been widely used to evaluate the performance of recommender systems. We use the two largest categories, i.e., Books and Movies \& TV, for the cross-domain CTR prediction task. The source domain is the book and the target domain is the movie.
We only keep users that have at least 5 ratings on items with metadata in each domain. We convert the ratings of 4-5 as label 1 and others as label 0. To simulate the industrial practice of CTR prediction (i.e., to predict the future CTR but not the past), we sort user logs in chronological order and take out the last rating of each user to form the test set, the second last rating to form the validation set and others to form the initial training set. The features used include 1) user features: user ID, and 2) book / movie features: item ID, brand, title, main category and categories.

\begin{table}[!t]
\setlength{\tabcolsep}{2pt}
\renewcommand{\arraystretch}{1.2}
\caption{Statistics of experimental datasets. (fts. - features, ini. - initial, insts. - instances, val. - validation, avg. - average)}
\vskip -8pt
\label{tab_stat}
\centering
\begin{tabular}{|l|p{2.4cm}|c c|}
\hline
\multirow{7}{*}{Company} & & Source: News & Target: Ad \\
\cline{2-4}
& \# Fields & User: 5, News: 18 & User: 5, Ad: 13 \\
& \# Unique fts. & \multicolumn{2}{c|}{10,868,554} \\
& \# Ini. train insts. & 53,776,761 & 11,947,267 \\
& \# Val. insts. & 8,834,570 & 1,995,980 \\
& \# Test insts. & 8,525,115 & 1,889,092 \\
& Max/Avg. \# clicked (*) per target ad & 25 / 7.37 & 5 / 1.10 \\
\hline
\hline
\multirow{9}{*}{Amazon}  & & Source: Book & Target: Movie \\
\cline{2-4}
& \# Fields & User: 1, Book: 5 & User: 1 , Movie: 5 \\
& \# Shared users & \multicolumn{2}{c|}{20,479} \\
& \# Unique fts. & \multicolumn{2}{c|}{841,927} \\
& \# Ini. train insts. & 794,048 & 328,005 \\
& \# Val. insts. & 20,479 & 20,479 \\
& \# Test insts. & 20,479 & 20,479 \\
& Max/Avg. \# clicked (*) per target movie & 20 / 9.82 & 10 / 6.69 \\
\hline
\end{tabular}
\vskip -8pt
\end{table}

\subsection{Methods in Comparison}
We compare both single-domain and cross-domain methods.
Existing cross-domain methods are mainly proposed for cross-domain recommendation and we extend them for cross-domain CTR prediction when necessary (e.g., to include attribute features rather than only IDs and to change the loss function).

\subsubsection{Single-Domain Methods}
\begin{enumerate}
\item \textbf{LR}. Logistic Regression \cite{richardson2007predicting}. It is a generalized linear model.
\item \textbf{FM}. Factorization Machine \cite{rendle2010factorization}. It models both first-order feature importance and second-order feature interactions.
\item \textbf{DNN}. Deep Neural Network \cite{cheng2016wide}. It contains an embedding layer, several FC layers and an output layer.
\item \textbf{Wide\&Deep}. Wide\&Deep model \cite{cheng2016wide}. It combines LR (wide part) and DNN (deep part).
\item \textbf{DeepFM}. DeepFM model \cite{guo2017deepfm}. It combines FM (wide part) and DNN (deep part).
\item \textbf{DeepMP}. Deep Matching and Prediction model \cite{ouyang2019representation}. It learns more representative feature embeddings for CTR prediction.
\item \textbf{DIN}. Deep Interest Network model \cite{zhou2018deep}. It models dynamic user interest based on historical behavior for CTR prediction.
\item \textbf{DSTN}. Deep Spatio-Temporal Network model \cite{ouyang2019deep}. It exploits spatial and temporal auxiliary information (i.e., contextual, clicked and unclicked ads) for CTR prediction.
\end{enumerate}

\subsubsection{Cross-Domain Methods}
\begin{enumerate}
\item \textbf{CCCFNet}. Cross-domain Content-boosted Collaborative Filtering Network \cite{lian2017cccfnet}. It is a factorization framework that ties collaborative filtering (CF) and content-based filtering. It relates to the neural network (NN) because latent factors in CF is equivalent to embedding vectors in NN.
\item \textbf{MV-DNN}. Multi-View DNN model \cite{elkahky2015multi}. It extends the Deep Structured Semantic Model (DSSM) \cite{huang2013learning} and has a multi-tower matching structure.
\item \textbf{MLP++}. MLP++ model \cite{hu2018conet}. It combines two MLPs with shared user embeddings across domains.
\item \textbf{CoNet}. Collaborative cross Network \cite{hu2018conet}. It adds cross connection units on MLP++ to enable dual knowledge transfer.
\item \textbf{MiNet}. Mixed Interest Network proposed in this paper.
\end{enumerate}

\newcommand{\tabincell}[2]{\begin{tabular}{@{}#1@{}}#2\end{tabular}}

\begin{table}[!t]
\setlength{\tabcolsep}{3pt}
\renewcommand{\arraystretch}{1.2}
\caption{Test AUC and Logloss. * indicates the statistical significance for $p<=0.01$ compared with the best baseline method based on the paired t-test.}
\vskip -10pt
\label{tab_auc}
\centering
\begin{tabular}{|l|l|c|c||c|c|}
\hline
\multicolumn{2}{|c}{} & \multicolumn{2}{|c||}{\textbf{Company}} & \multicolumn{2}{|c|}{\textbf{Amazon}} \\
\hline
\multicolumn{2}{|c|}{\textbf{Model}} & AUC & Logloss & AUC & Logloss \\
\hline
\multirow{7}{*}{\tabincell{c}{Single-\\domain}} & LR & 0.6678 & 0.5147 &0.7173 & 0.4977 \\
& FM & 0.6713 & 0.5133 & 0.7380 & 0.4483 \\
& DNN & 0.7167 & 0.4884 & 0.7688 & 0.4397 \\
& Wide\&Deep & 0.7178 & 0.4879 & 0.7699 & 0.4389 \\
& DeepFM & 0.7149 & 0.4898 & 0.7689 & 0.4406 \\
& DeepMP & 0.7215 & 0.4860 & 0.7714 & 0.4382 \\
& DIN & 0.7241 & 0.4837 & 0.7704 & 0.4393 \\
& DSTN & 0.7268 & 0.4822 & 0.7720 & 0.4296 \\
\hline
\multirow{5}{*}{\tabincell{c}{Cross-\\domain}} & CCCFNet & 0.6967 & 0.5162 & 0.7518 & 0.4470 \\
& MV-DNN & 0.7184 & 0.4875 & 0.7814 & 0.4298 \\
& MLP++ & 0.7192 & 0.4878 & 0.7813 & 0.4306\\
& CoNet & 0.7175 & 0.4882 & 0.7791 & 0.4389 \\
& MiNet & \textbf{0.7326}* & \textbf{0.4784}* & \textbf{0.7855}* & \textbf{0.4254}* \\
\hline
\end{tabular}
\vskip -8pt
\end{table}

\subsection{Parameter Settings}
We set the dimension of the embedding vectors for each feature as $D=10$. 
We set $C=10$ and the number of FC layers in neural network-based models as $L = 2$. The dimensions are [512, 256] for the Company dataset and [256, 128] for the Amazon dataset. We set $D_h = 128$ for the Company dataset and $D_h = 64$ for the Amazon dataset. The batch sizes for the source and the target domains are set to (512, 128) for the Company dataset and (64, 32) for the Amazon dataset.
All the methods are implemented in Tensorflow and optimized by the Adagrad algorithm \cite{duchi2011adaptive}. We run each method 5 times and report the average results.

\subsection{Evaluation Metrics}
\begin{enumerate}
\item \textbf{AUC}: the Area Under the ROC Curve over the test set (target domain). It is a widely used metric for CTR prediction. It reflects the probability that a model ranks a randomly chosen positive instance higher than a randomly chosen negative instance. The larger the better. A small improvement in AUC is likely to lead to a significant increase in online CTR.
\item \textbf{RelaImpr}: RelaImpr is introduced in \cite{yan2014coupled} to measure the relative improvement of a target model over a base model. Because the AUC of a random model is 0.5, it is defined as: $RelaImpr = \left( \frac{\textrm{AUC(target model)} - 0.5}{\textrm{AUC(base model)} - 0.5} -1 \right) \times 100\%$.
\item \textbf{Logloss}: the value of Eq. (\ref{loss_t}) over the test set (target domain). The smaller the better.
\end{enumerate}

\begin{figure}[!t]
\centering
\includegraphics[width=0.46\textwidth, trim = 10 10 5 10, clip]{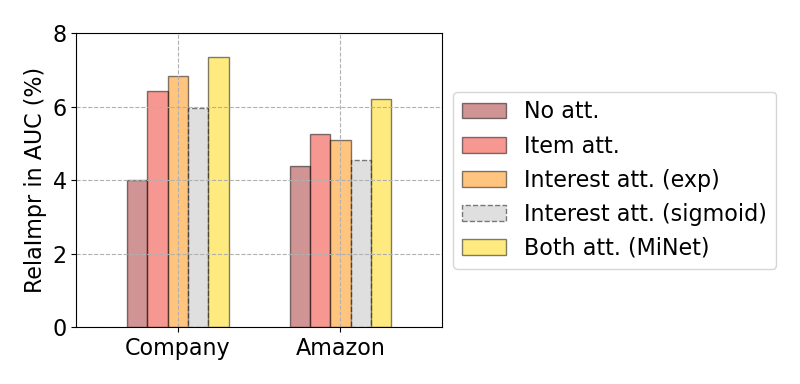}
\vskip -8pt
\caption{Effect of level of attention (att.). Base model: DNN.}
\vskip -8pt
\label{fig_att}
\end{figure}

\subsection{Effectiveness}
Table \ref{tab_auc} lists the AUC and Logloss values of different methods. It is observed that in terms of AUC, shallow models such as LR and FM perform worse than deep models. FM performs better than LR because it further models second-order feature interactions.
Wide\&Deep achieves higher AUC than LR and DNN, showing that combining LR and DNN can improve the prediction performance. Among the single-domain methods, DSTN performs best. It is because DSTN jointly considers various spatial and temporal factors that could impact the CTR of the target ad.

As to the cross-domain methods, CCCFNet outperforms LR and FM, showing that using cross-domain data can lead to improved performance. CCCFNet performs worse than other cross-domain methods because it compresses all the attribute features. MV-DNN performs similarly as MLP++. They both achieve knowledge transfer through embedding sharing. CoNet introduces cross connection units on MLP++ to enable dual knowledge transfer across domains. However, this also introduces higher complexity and random noise. CCCFNet, MV-DNN, MLP++ and CoNet mainly consider long-term user interest. In contrast, our proposed MiNet considers not only long-term user interest, but also short-term interest in both domains. With proper combination of these different interest signals, MiNet outperforms other methods significantly.

\subsection{Ablation Study: Level of Attention}
In this section, we examine the effect of the level of attentions in MiNet.
In particular, we examine the following settings: 1) No attention, 2) Only item-level attention, 3) Only interest-level attention [with the exponential activation function as proposed in Eq. (\ref{att_int})], 4) Only interest-level attention but with the sigmoid activation function [i.e., replacing $\exp$ by sigmoid in Eq. (\ref{att_int})], and 5) Both attentions.
It is observed in Figure \ref{fig_att} that ``No attention'' performs worst. This is because useful signals could be easily buried in noise without distilling. Either item-level attention or interest-level attention can improve the AUC, and the use of both levels of attentions results in the highest AUC. Moreover, ``Interest attention (sigmoid)'' has much worse performance than ``Interest attention (exp)''. This is because improper activation function cannot effectively tackle the dimension discrepancy problem. These results demonstrate the effectiveness of the proposed hierarchical attention mechanism.

\subsection{Ablation Study: Attention Weights}
\begin{figure}[!t]
\centering
\includegraphics[width=0.48\textwidth, trim = 0 0 0 00, clip]{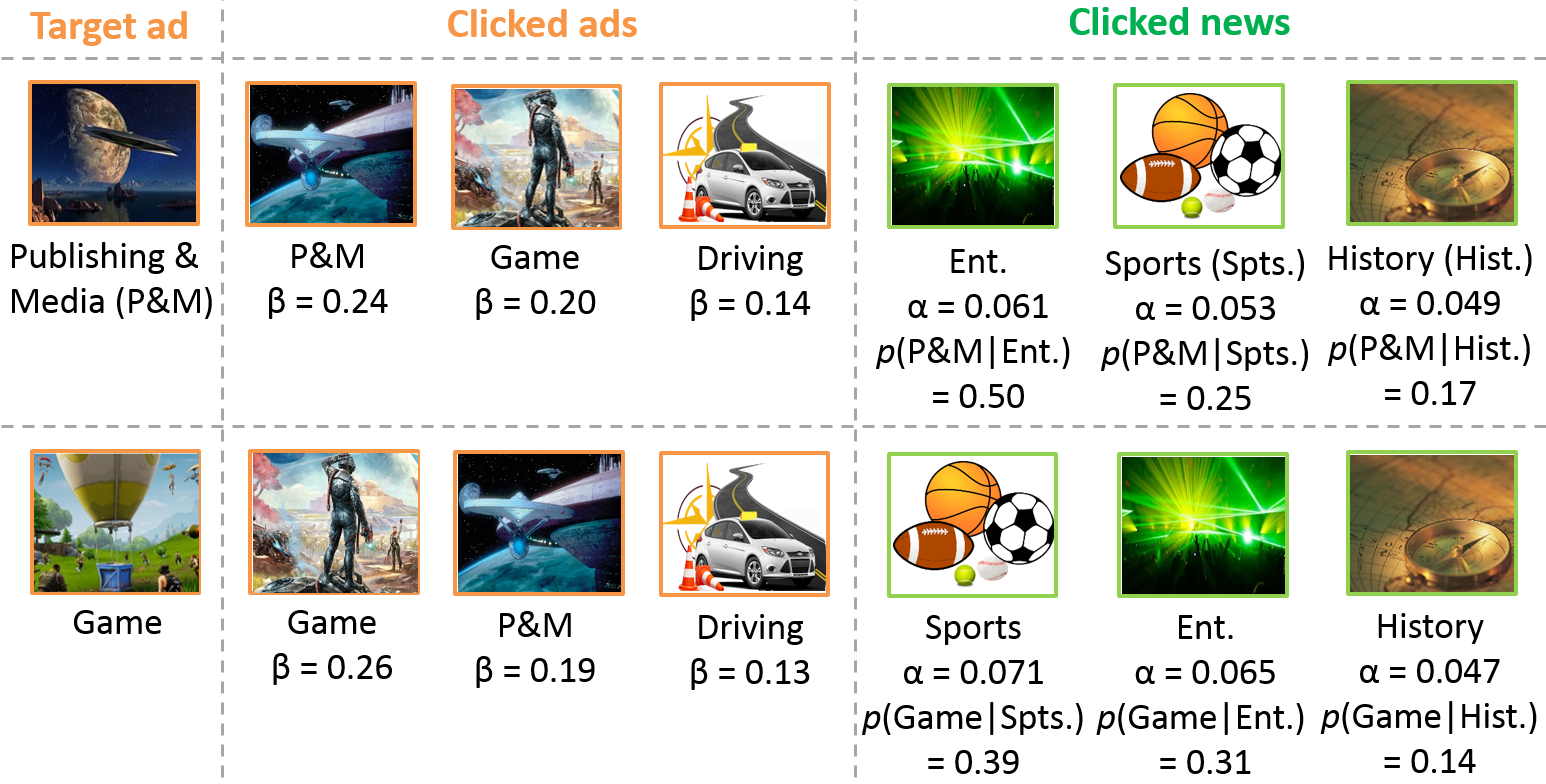}
\vskip -8pt
\caption{Item-level attention weights for an active user (two \emph{different} target ads with the \emph{same} set of clicked ads and news in the category granularity). Pictures are only for illustration. Attention weights do dynamically change w.r.t. different target ads. (Ent. - Entertainment)}
\vskip -8pt
\label{fig_wgt}
\end{figure}

In this section, we examine the item-level attention weights in MiNet and check whether they can capture informative signals.
Figure \ref{fig_wgt} shows two \emph{different} target ads with the \emph{same} set of clicked ads and clicked news for an active user in the Company dataset. For the privacy issue, we only present ads and news at the category granularity. Since ads are in the same domain, it is relatively easy to judge the relevance between a clicked ad and the target ad. Nevertheless, as ads and news are in different domains, it is hard to judge their relevance. We thus calculate the probability $p(Ad | News)$ based on the user's behavior logs.

It is observed in Figure \ref{fig_wgt} that when the target ad is Publishing \& Media (P\&M), the clicked ad of P\&M has the highest weight and the clicked news of Entertainment has the highest weight; but when the target ad is Game, the clicked ad of Game has the highest weight and the clicked news of Sports has the highest weight. These results show that the item-level attentions do dynamically capture more important information for different target ads. It is also observed that the model can learn certain correlation between a piece of clicked news and the target ad. News with a higher indication probability usually receives a higher attention weight.

\subsection{Ablation Study: Effect of Modeling Different Types of User Interest}
\begin{figure}[!t]
\centering
\includegraphics[width=0.46\textwidth, trim = 10 10 5 10, clip]{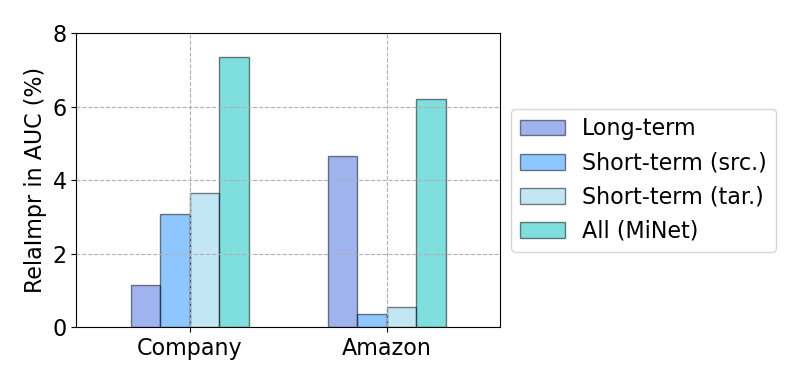}
\vskip -8pt
\caption{Effect of modeling long- and short-term interest. Base model: DNN. (src. - source domain, tar. - target domain)}
\vskip -4pt
\label{fig_interest}
\end{figure}

In this section, we examine the effect of modeling different types of user interest in MiNet.
We observe quite different phenomena on the two datasets in Figure \ref{fig_interest}. On the Company dataset, modeling short-term interest can lead to much higher AUC than modeling long-term interest, showing that recent behaviors are quite informative in online advertising. In contrast, on the Amazon dataset, modeling long-term interest results in much higher AUC. It is because the Amazon dataset is an e-commerce dataset rather than advertising and the nature of ratings are different from that of clicks. Nevertheless, when all these aspects are jointly considered in MiNet, we obtain the highest AUC, showing that different types of interest can complement each other and joint modeling can lead to the best and more robust performance.

\subsection{Online Deployment} \label{online}
\begin{figure}[!t]
\centering
\includegraphics[width=0.48\textwidth, trim = 0 0 0 0, clip]{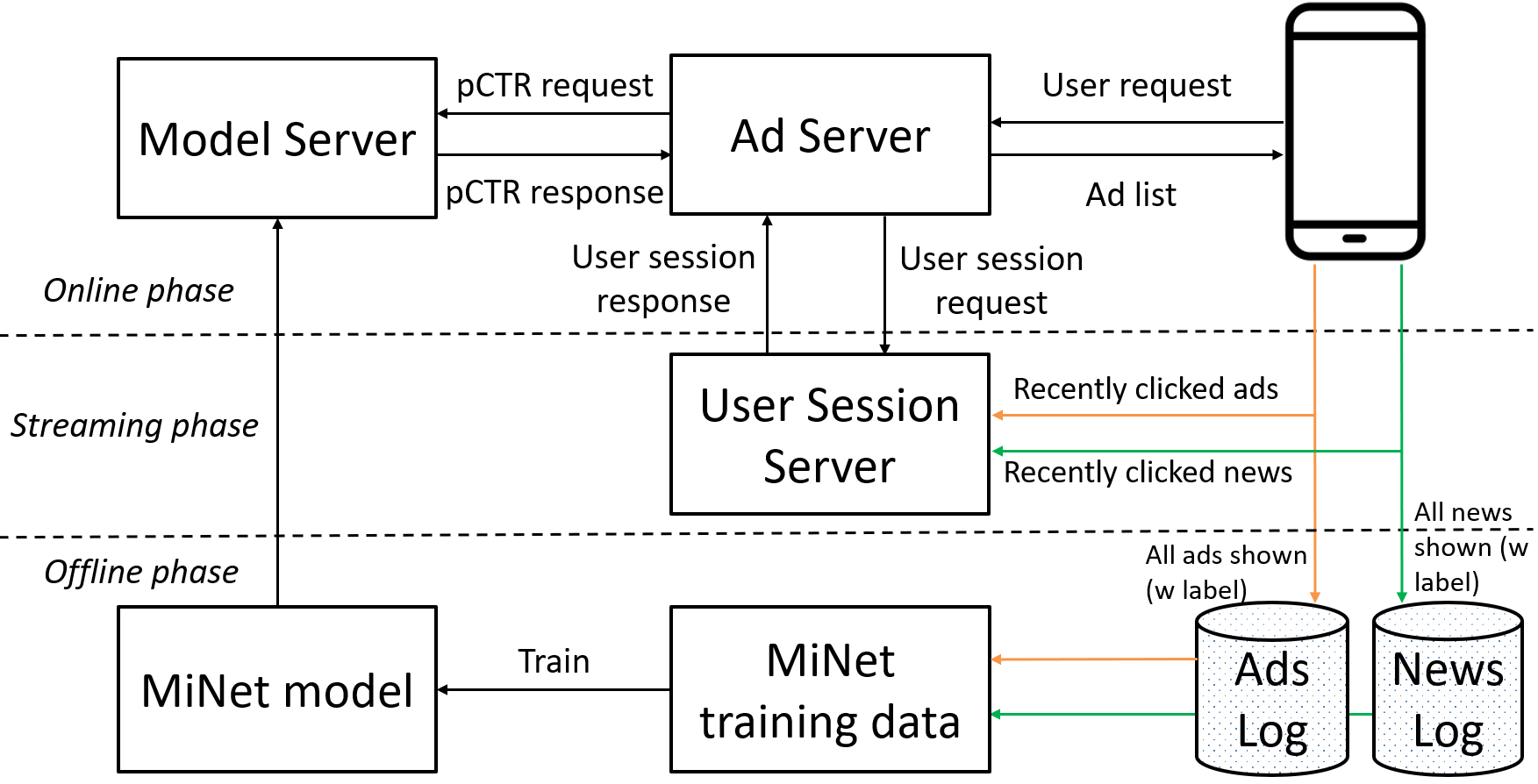}
\vskip -8pt
\caption{Architecture of the ad serving system with MiNet.}
\vskip -8pt
\label{fig_system}
\end{figure}

We deployed MiNet in UC Toutiao, where the ad serving system architecture is shown in Figure \ref{fig_system}.
We conducted online experiments in an A/B test framework over two weeks during Dec. 2019 - Jan. 2020, where the base serving model is DSTN \cite{ouyang2019deep}.
Our online evaluation metric is the real CTR, which is defined as the number of clicks over the number of ad impressions. A larger online CTR indicates the enhanced effectiveness of a CTR prediction model. The online A/B test shows that MiNet leads to an increase of online CTR of 4.12\% compared with DSTN. This result demonstrates the effectiveness of MiNet in practical CTR prediction tasks. After the A/B test, MiNet serves the main ad traffic in UC Toutiao.

\section{Related Work}
\textbf{CTR prediction.}
Existing works mainly address the single-domain CTR prediction problem. They model aspects such as 1) feature interaction (e.g., FM \cite{rendle2010factorization} and DeepFM \cite{guo2017deepfm}), 2) feature embeddings (e.g., DeepMP \cite{ouyang2019representation}), 3) user historical behavior (e.g., DIN \cite{zhou2018deep} and DSTN \cite{ouyang2019deep}) and 4) contextual information (e.g., DSTN \cite{ouyang2019deep}).

As generalized linear models such as Logistic Regression (LR) \cite{richardson2007predicting} and Follow-The-Regularized-Leader (FTRL) \cite{mcmahan2013ad} lack the ability to learn sophisticated feature interactions \cite{chapelle2015simple}, Factorization Machine (FM) \cite{rendle2010factorization, blondel2016higher} is proposed to address this limitation. Field-aware FM \cite{juan2016field} and Field-weighted FM \cite{pan2018field} further improve FM by considering the impact of the field that a feature belongs to. In recent years, neural network models such as Deep Neural Network (DNN) and Product-based Neural Network (PNN) \cite{qu2016product} are proposed to automatically learn feature representations and high-order feature interactions \cite{covington2016deep,wang2017deep}. Some models such as Wide\&Deep \cite{cheng2016wide}, DeepFM \cite{guo2017deepfm} and Neural Factorization Machine (NFM) \cite{he2017neural} combine a shallow model and a deep model to capture both low- and high-order feature interactions. Deep Matching and Prediction (DeepMP) model \cite{ouyang2019representation} combines two subnets to learn more representative feature embeddings for CTR prediction.

Deep Interest Network (DIN) \cite{zhou2018deep} and Deep Interest Evolution Network (DIEN) \cite{zhou2019deep} model user interest based on historical click behavior. Xiong et al. \cite{xiong2012relational} and Yin et al. \cite{yin2014exploiting} consider various contextual factors such as ad interaction, ad depth and query diversity. Deep Spatio-Temporal Network (DSTN) \cite{ouyang2019deep} jointly exploits contextual ads, clicked ads and unclicked ads for CTR prediction.

\textbf{Cross-domain recommendation.}
Cross-domain recommendation aims at improving the recommendation performance of the target domain by transferring knowledge from source domains.
These methods can be broadly classified into three categories: 1) collaborative \cite{singh2008relational,man2017cross}, 2) content-based \cite{elkahky2015multi} and 3) hybrid \cite{lian2017cccfnet}.

Collaborative methods utilize interaction data (e.g., ratings) across domains. For example, Ajit et al. \cite{singh2008relational} propose Collective Matrix Factorization (CMF) which assumes a common global user factor matrix and factorizes matrices from multiple domains simultaneously.
Gao et al. \cite{gao2019cross} propose the Neural Attentive Transfer Recommendation (NATR) for cross-domain recommendation without sharing user-relevant data. Hu et al. \cite{hu2018conet} propose the Collaborative cross Network (CoNet) which enables dual knowledge transfer across domains by cross connections.
Content-based methods utilize attributes of users or items. For example, Elkahky et al. \cite{elkahky2015multi} transform the user profile and item attributes to dense vectors and match them in a latent space. Zhang et al. \cite{zhang2016collaborative} utilize textual, structure and visual knowledge of items as auxiliary information to aid the learning of item embeddings.
Hybrid methods combine interaction data and attribute data. For example, Lian et al. \cite{lian2017cccfnet} combine collaborative filtering and content-based filtering in a unified framework.

Differently, in this paper, we address the cross-domain CTR prediction problem. We model three types of user interest and fuse them adaptively in a neural network framework.

\section{Conclusion}
In this paper, we address the cross-domain CTR prediction problem for online advertising. We propose a new method called the Mixed Interest Network (MiNet) which models three types of user interest: long-term interest across domains, short-term interest from the source domain and short-term interest in the target domain. MiNet contains two levels of attentions, where the item-level attention can dynamically distill useful information from recently clicked news / ads, and the interest-level attention can adaptively adjust the importance of different user interest signals. Offline experiments demonstrate the effectiveness of the modeling of three types of user interest and the use of hierarchical attentions. Online A/B test results also demonstrate the effectiveness of the model in real CTR prediction tasks in online advertising.

\bibliographystyle{ACM-Reference-Format}
\bibliography{ref}

\end{document}